\documentclass[sn-chicago]{sn-jnl}

\jyear{2022}%

\theoremstyle{thmstyleone}%
\theoremstyle{thmstyletwo}%

\theoremstyle{thmstylethree}%
\newcounter{example}[section]
\newenvironment{example}[1][]{\refstepcounter{example}\par\medskip
    \noindent\textbf{Example~\theexample: #1} \rmfamily}{\medskip}

\newcommand\posscite[1]{\citeauthor{#1}'s (\citeyear{#1})}
\usepackage{capt-of}
\usepackage{xcolor}

\begin{document}

\title[Acknowledgement Texts in WoS]{A Comprehensive Analysis of Acknowledgement Texts in Web of Science: \\ a case study on four scientific domains}


\author*[1]{\fnm{Nina} \sur{Smirnova}}\email{ninasmirnova@web.de}

\author*[1]{\fnm{Philipp} \sur{Mayr}}\email{philipp.mayr@gesis.org}


\affil[1]{\orgname{GESIS –- Leibniz Institute for the Social Sciences}, \orgaddress{\street{Unter Sachsenhausen 6-8}, \city{Cologne}, \postcode{50667}, \country{Germany}}}




\abstract{Analysis of acknowledgments is particularly interesting as acknowledgments may give information not only about funding, but they are also able to reveal hidden contributions to authorship and the researcher’s collaboration patterns, context in which research was conducted, and specific aspects of the academic work. The focus of the present research is the analysis of a large sample of acknowledgement texts indexed in the Web of Science (WoS) Core Collection. Record types “article” and “review” from four different scientific domains, namely social sciences, economics, oceanography and computer science, published from 2014 to 2019 in a scientific journal in English were considered. Six types of acknowledged entities, i.e., funding agency, grant number, individuals, university, corporation and miscellaneous, were extracted from the acknowledgement texts using a Named Entity Recognition (NER) tagger and subsequently examined. A general analysis of the acknowledgement texts showed that indexing of funding information in WoS is incomplete. 
The analysis of the automatically extracted entities revealed differences and distinct patterns in the distribution of acknowledged entities of different types between different scientific domains. A strong association was found between acknowledged entity and scientific domain, and acknowledged entity and entity type. 
Only negligible correlation was found between the number of citations and the number of acknowledged entities.
Generally, the number of words in the acknowledgement texts positively correlates with the number of acknowledged funding organizations, universities, individuals and miscellaneous entities. At the same time, acknowledgement texts with the larger number of sentences have more acknowledged individuals and miscellaneous categories. 
}

\keywords{Acknowledgements, Web of Science, Acknowledged Entities, Named Entity Recognition}



\maketitle

\section{Introduction}\label{sec:intro}
 
Acknowledgments in scientific papers are short texts where the author(s) express \textit{"gratitude towards different types of support received during the research process" }\citep{alvarez_2021}. Kassierer and Angell  see acknowledgements as an instrument to \textit{“identify those who made special intellectual or technical contribution to a study that are not sufficient to qualify them for authorship” }\citep[p.~1511]{kassirer_authorship_1991}. \cite{cronin_praxis_1995} ascribe an acknowledgment alongside authorship and citedness to measures of a researcher's scholarly performance. Together, they build \textit{“The Reward Triangle”}: a feature that reflects the researcher’s productivity and impact \citep[p.~173]{cronin_praxis_1995}. Diaz-Faes and Bordons argue, \textit{“citations provide a measure of the underlying intellectual influence and foundations of research output”} \citep[p.~577]{diaz-faes_making_2017}. 

Acknowledgments as a rule contain information about technical, instrumental and financial support together with intellectual and conceptual support \citep{diaz-faes_making_2017,giles_who_2004}. The latter fall into the category \textit{“peer interactive communication”} (PIC) \citep{mccain_2018}. \cite{giles_who_2004} argue that PIC acknowledgments, in the same way as citations, may be used as a metric to measure an individual’s intellectual contribution to scientific work. Acknowledgements of financial support are interesting in terms of evaluating the influence of funding agencies on academic research. Acknowledgments of technical and instrumental support may reveal \textit{“indirect contributions of research laboratories and universities to research activities” }\citep[p.~17599]{giles_who_2004}. 
Thus, acknowledgements not only give information about funding, but they are also able to reveal hidden contributions to authorship and the researcher’s collaboration patterns, context in which research was conducted, and specific aspects of the academic work. 

In this paper, acknowledgement texts from four scientific domains (economics, social sciences, oceanography and computer science) were gathered from the Web of Science (WoS) database containing the collection of funding acknowledgments. The following acknowledged entities were extracted from the acknowledgement texts and distinguished into six categories: funding agencies (FUND), grant numbers (GRNB), corporations (COR), universities (UNI), individuals (IND), and miscellaneous (MISC), as Figure~\ref{fig:example_ackn} demonstrates. An \textit{acknowledged entity} is an object of acknowledgment. Acknowledged entities could be names and surnames of individuals (also including abbreviations of the names), names of institutions and organizations, and numbers and identification numbers of grants, as Figure~\ref{fig:example_ackn} shows. A comprehensive analysis was performed on the acknowledgement texts and extracted acknowledged entities.

Most of the previous works on acknowledgement analysis were limited by the manual evaluation of data and therefore by the amount of processed data \citep{giles_who_2004, paul_hus_all_2017,paul_hus_des_2019, Mccain2017}. Furthermore, to our knowledge, previous works on automatic acknowledgment analysis were mostly concerned with the extraction and analysis of funding organizations and grant numbers \citep{alexandera_this_2021,kayal-etal-2017-tagging} or classification of acknowledgement texts \citep{song_kang_timakum_zhang}. 
Moreover, large bibliographic databases such as Web of Science (WoS)\footnote{\url{http://wokinfo.com/products_tools/multidisciplinary/webofscience/fundingsearch/}} and Scopus selectively index only funding information, i.e., names of funding organizations and grant identification numbers. Therefore, to our knowledge, there are no previous works on acknowledgement analysis, which examines a large quantity of acknowledged entities other than funding organisations and grant numbers \citep{smirnova_mayr_2022}.
In the present paper, we have used a NER model, specifically trained to extract six types of entities described above, and conducted an extensive analysis of the great amount of the obtained acknowledged entities. 

Acknowledged entities are potentially a great tool for the analysis of different aspects of the scientific society. Thus, \cite{thomer_weber_2014} argue that using named entities can benefit the process of manual document classification and evaluation of the data. \cite{petrovich_2022} used the Acknowledged Entities Network (along with other acknowledgement related networks), containing funding agencies, congresses and conferences, institutions and persons, to develop a method for mapping scientiﬁc and scholarly social networks.



\paragraph{Research questions}\label{subsec:questions}
In the present paper, we will investigate two broad research directions: the general acknowledgment trends among different disciplines and the relationships between different variables: entity types, acknowledged entities, scientific domains, length of the acknowledgement text, number of citations. 
 
The first research direction will focus on the research question (RQ) 1:
\begin{itemize}
    \item \textbf{RQ1:} What are the general acknowledgment trends among different disciplines, i.e., which are the most acknowledged organizations, individuals, universities, or corporations? Do they vary between disciplines? How do acknowledgment patterns differ among disciplines and what do they have in common? Does the length of the acknowledgment vary between scientific domains?
 \end{itemize}
 
 The second research direction will address the research questions 2 and 3:
 \begin{itemize}
    \item \textbf{RQ2:} Are there correlations between the acknowledged entities and the scientific domains?
    \item \textbf{RQ3:} Do highly-cited papers have more funding entities acknowledged than paper with fewer citations?
 \end{itemize}
 
Previous works in acknowledgment analysis show that acknowledgement patterns diverge in different scientific domains \citep{paul_hus_all_2017,diaz-faes_making_2017, baccini_petrovich_2021}.
In this paper, we expect to find correlations between the scientific domain and the type of acknowledged entity. 
We assume (similar to \cite{paul_hus_all_2017,diaz-faes_making_2017}) that social sciences and economics will show the highest number of acknowledgments of peer-interactive communication (acknowledged individuals). Computer science and oceanography are expected to have a high percentage of acknowledged entities carrying information about funders (names of funding organizations and projects, grant numbers) and affiliated universities. Generally, more entities are expected to fall into the FUND and GRNB categories, as WoS stores only acknowledgements containing funding information. It is also expected to find a correlation between acknowledged entity and label, i.e. precise entities will have a precise label, and consequently, there should be a correlation between acknowledged entity and scientific domain. 

 
 
\section{Background and Related Work}\label{sec:background}

Currently, there are several major bibliographic databases \citep{singh2021}. The following major bibliographic databases can be distinguished: Web of Science (WoS), SCOPUS and Dimensions.
WoS provides subscription-based access to publisher-independent global citation databases\footnote{\url{https://clarivate.com/webofsciencegroup/solutions/web-of-science/}}. WoS contains publications from different scientific fields, and the WoS Core Collection consists of a number of databases.

\begin{figure}[h!]
\centering
  \includegraphics[width=1\textwidth]{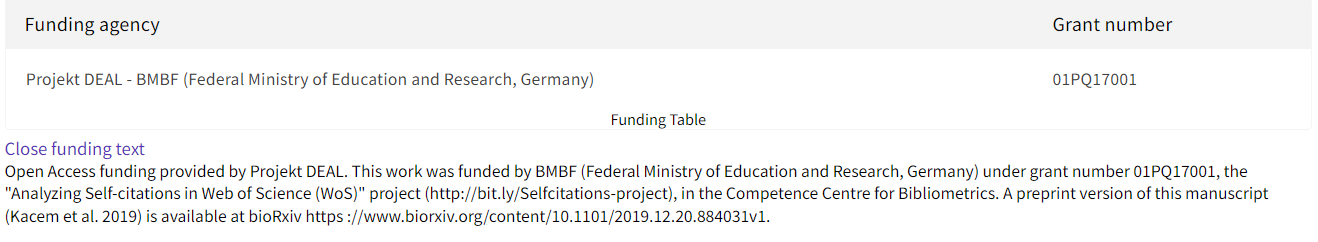}
  \caption{An example of funding information indexed in WoS.}
  \label{fig:example_wos}
\end{figure}

From 2008 on, WoS started indexing funding information (mainly funding agencies and grant numbers) in its databases\footnote{\url{http://wokinfo.com/products_tools/multidisciplinary/webofscience/fundingsearch/}}. WoS uses information from different funding reporting systems such as researchfish\footnote{\url{https://mrc.ukri.org/funding/guidance-for-mrc-award-holders/researchfish/}}, Medline\footnote{\url{https://www.nlm.nih.gov/bsd/funding_support.html}} and others. The funding information indexed in WoS is distinguished into three different fields: "Funding Text" (FT), "Funding Agency" (FO) and "Grant Number" (FG). The FT-field contains the full text of the acknowledgement, the FO-field contains the names of organizations acknowledged for their funding contribution, and the FG-field contains grant numbers affiliated with the funding organizations identified in the FO-field \citep[pp.~170-171]{paul_hus_all_2016}. Figure \ref{fig:example_wos} demonstrates an example of funding information indexed in WoS. 

Research in the field of acknowledgements analysis has been carried out since the 1970s. The first typology of acknowledgements was proposed by \cite{mackintosh_1972} \citep[as cited in][]{Cronin1995TheSC}. The acknowledgement texts found in American Sociological Review were examined and distinguished into three categories: facilities, access to data, help of individuals. \cite{mccain_1991} proposed a five-level typology based on the analysis of acknowledgement text from the genetics domain: research related information, secondary access to research-related information, specific research-related communication, general peer-communication, technical or clerical support. \cite{Cronin1995TheSC} distinguished acknowledgements to the three broad categories: resource related, procedure-related and concept related. \cite{mejia_using_2018} analysed acknowledgement texts from the WoS Core Collection from the robotic research field. Aim of the analysis was to characterize funding organizations by the types of the sponsored research field. \cite{mejia_using_2018} developed a four-level classification of research fields: change maker, incremental, breakthrough and matured. Incremental and matured research fields showed the highest number of acknowledged funding organizations. The authors claimed that the classification of funding agencies may benefit the development of funding strategies of policy makers and funding organizations.  
\cite{paul_hus_des_2019} created an extended classification of acknowledgments stored in Web of Science (WoS). About four thousand acknowledgment sentences were manually coded into 13 categories. The analysis of coded data reveals three distinct axes:  the contribution, the disclaimers, and the authorial voice \citep[p.~5]{paul_hus_des_2019}.

\cite{lewison_publications_1994} analysed WoS SCI papers funded by the Biotechnology Action Programme (BAP) in the period 1986-1993. The aim of the analysis was to investigate the level of multi-nationality of the publications compared to other European Community biotechnology programs.
\cite{tollison_labalnd_2003} examined asterisk footnotes from three major journals of economics, to investigate acknowledging behaviour in economics. Analysis shows that the most famous economists are acknowledged the most. 
\cite{giles_who_2004} proposed a method for the automated extraction of acknowledged entities from the text and analysis of acknowledgment texts.  
Research papers from the computer science domain deposited in the CiteSeer digital library were used as the data source. Extracted acknowledged entities were linked to the source articles and to the data from CiteSer’s citation index for further analysis. Based on manual analysis of the most acknowledged entities, Giles and Councill developed a four-level classification:  funding agencies, corporations, universities, and individuals. 
\cite{wang_funding_2011} investigated the connection between research funding and the development of science and technology. Articles from the field of nanotechnology published in the period 2008-2009 were included in the study. The analysis showed that \textit{"most of nanotechnology funding is nationally-oriented"}\textit{} \citep{wang_funding_2011}. International collaborations are mainly indicated by the funding of individual international researchers. 
\cite{rigby_horns_2014} studied WoS collection of acknowledgement texts containing funding from European Molecular Biology Organisation or the Human Frontier Science Program published in the period from 2008-2012. Analysis of the patterns indicating over-finance of the particular project, i.g., projects funded by more than one funding organization in a similar area, has greater citation impact. 
\cite{paul_hus_all_2016} made an overview on the funding data stored in WoS. Data from all WoS Core Collection databases was used for the analysis: Science Citation Index Expanded (SCIE), Social Science Citation Index (SSCI), and Arts \& Humanities Citation Index (AHCI). SCIE provides a multidisciplinary search in scientific journals, SSCI covers journals in social sciences disciplines, and AHCI contains publications in the fields of arts and humanities\footnote{\url{https://clarivate.com/webofsciencegroup/solutions/web-of-science-core-collection/}}. 
The aim of the research was to help other researchers understand the potential and limitations of funding information. The analysis was performed on all publications included in WoS for the period from 2005 to 2015 and comprised more than 43 million documents.  The analysis of distribution of funding texts by year showed that the full coverage of funding text data starts from the year 2009, therefore, the collection of funding texts published before 2009 is incomplete. Funding information is not included in every WoS core collection database. SCIE is mostly covered in terms of funding text data. Funding texts are also included in the SSCI starting from 2015, and the AHCI has no funding information indexation. Consequently, very little funding data was found for many humanities fields. Most funding information was found for publications written in English. Publications in other languages had an extremely low contribution of indexed acknowledgments texts. The main limitation of WoS data is that only acknowledgments containing funding information are included. It may bias the analysis of non-funding types of contribution. 
Acknowledgment patterns are argued to be domain-specific. \cite{paul_hus_all_2017} conducted an analysis of acknowledgments from research articles and reviews stored in WoS. The dataset was restricted by articles published in 2015. The aims of the study were to distinguish types of acknowledged contributions, and to find out how they vary by discipline. The results matched the observations of previous research \citep{paul_hus_all_2016}. The highest percentage of publications that included acknowledgments came from the biomedical and natural sciences, followed by clinical medicine. The lowest percentage of funding texts came from the social sciences. Thus, differences between frequencies of types of acknowledged entities across disciplines were found. For example, technical support was mostly acknowledged in publications in the natural sciences. Earth and space, professional fields, and social sciences tend to acknowledge peer contribution. Biomedical research showed a higher percentage of funding information. Acknowledgments in Biology are mainly focused on logistic and fieldwork-related tasks. \cite{diaz-faes_making_2017} observed the prevalence of PIC acknowledgments in the humanities domain.
\cite{tian_acknowledgement_2021} analysed an acknowledgement network to estimate the influence of the acknowledged person. The acknowledgement network for the analysis was created using the WoS collection of acknowledgement texts from the field of wind energy published between 2008 and 2010. The study showed that the acknowledged individual's centrality positively moderates the citation count. On the opposite authors' centrality in the collaboration network is negatively associated with the relationship between acknowledged individual's centrality and citation count of a paper.
\cite{baccini_petrovich_2021} studied acknowledging behaviour by analysing WoS articles from top-five journals of economics. Acknowledgements were assessed from two perspectives. According to the normative account, acknowledgements serve to recognize the contribution of other collaborators. According to the strategic account, acknowledging influential scholars increases the visibility and quality of the paper. The authors argue that both accounts \textit{"should be conceived as partial accounts of the various motivations behind the acknowledging behaviour of researchers"} \citep[p.~34]{baccini_petrovich_2021}. Their study showed the variations of acknowledgement behaviour among different fields. 
\cite{georg_rose_2015} examined informal collaborations in academic research. The authors developed a dataset containing 5000 acknowledgements from 6 high-ranking journals in economics. The dataset analysis revealed generational and gender differences in informal collaborations. The authors argue that \textit{"information derived from networks of informal collaboration allows us to predict academic impact of both researchers and papers even better than information from co-author networks."}

\section{Materials and Method}\label{sec:materials_method}

Acknowledgement texts from the WoS Core Collection
were used for the following analysis. 
As WoS contains millions of metadata records, the data chosen for the present study was restricted by year and scientific domain. Records from four different scientific domains published from 2014 to 2019 were considered: two domains from the social sciences (social sciences and economics), oceanography and computer science\footnote{The full list of the disciplines selected for the present study can be found in the Appendix \ref{app:wos_disc}.}. Four different domains were selected to analyse the interdisciplinary differences in acknowledgement behaviour. Social sciences and economics are closely associated domains, while oceanography is a branch of natural science and computer science is a broad technical scientific domain. In the MinAck project \footnote{\url{https://kalawinka.github.io/minack/}}, which is the context of this research, it is of interest to study the differences between closely and widely associated scientific domains.
For the present study, the acknowledgments corpus was restricted to approximately 200,000 entries. Therefore, approximately 50,000 records were taken randomly from each scientific domain which resulted in the total amount of records in the acknowledgments corpus of 196,875 entries. Table~\ref{tab:ackn_corpus_count} represents the total number of articles (research articles, reviews) in the corpus from each scientific domain restricted by these criteria.

\begin{table}[!h]
\begin{center}
\caption{Number of articles (research articles, reviews) selected for the analyzed acknowledgment corpus in each scientific domain}\label{tab:ackn_corpus_count}%
\begin{tabular}{p{4cm} p{5cm}}
\toprule
Scientific domain & Number of records in our acknowledgments corpus \\
\midrule
oceanography & 49,420 \\
economics &  49,487 \\
social sciences & 49,211 \\
computer science & 48,757 \\
\botrule
\end{tabular}
\end{center}
\end{table}

\begin{figure}[h!]
\centering
  \includegraphics[width=1\textwidth]{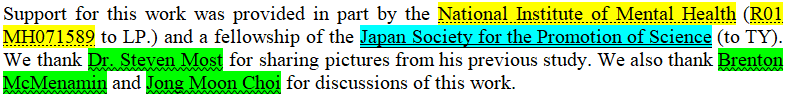}
  \caption{An example of WoS indexing problems. Entities indexed in WoS are marked yellow and with a dotted underline. Funding agencies not indexed in WoS are marked blue and with a double underline. Individuals (also not indexed in WoS) are marked green and with a wavy underline.}
  \label{fig:wos_missing_info}
\end{figure}

Extraction and classification of the acknowledged entities was performed using the Flair-NLP framework \citep{akbik_flair_2019}. The reason to use our own software is mainly contingent on problems with WoS funding information indexing: only information about funders is included; i.e., individuals are not indexed. Furthermore, the existing indexing of funding organizations is incomplete, as Figure~\ref{fig:wos_missing_info} demonstrates. Furthermore, indexed funding organizations are not divided into different entity types like universities, corporations, etc. 

\begin{figure}[h!]
\centering
  \includegraphics[width=0.7\textwidth]{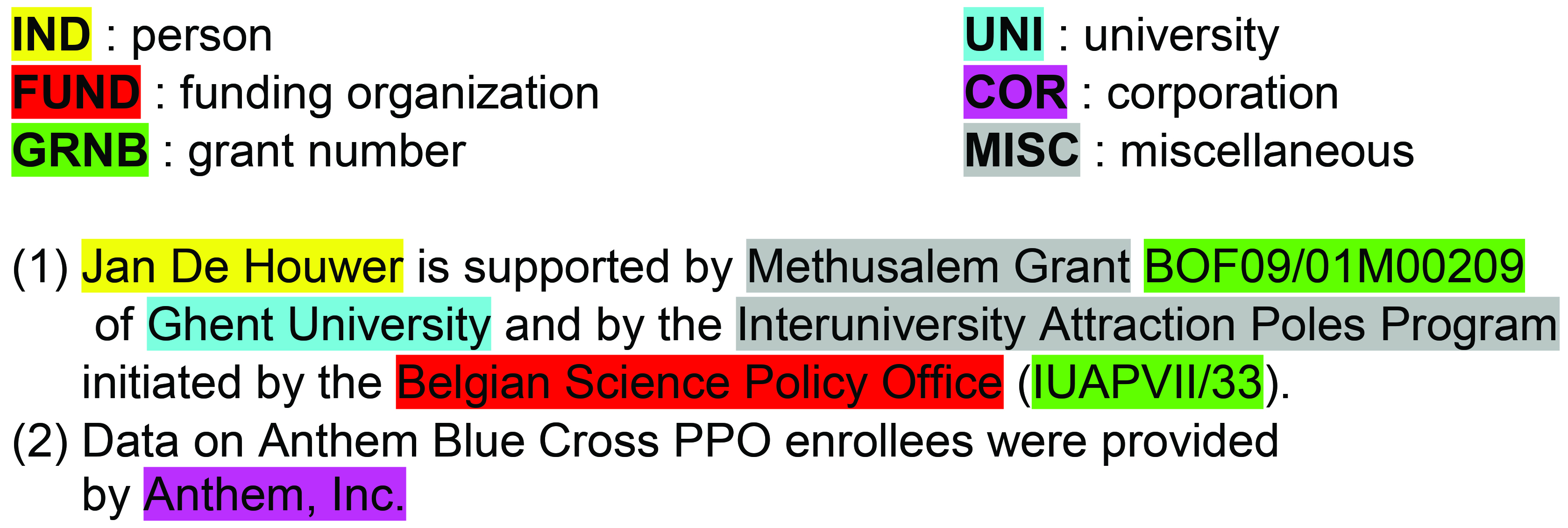}
  \caption{An example of acknowledged entities. Each entity type is marked with a distinct color.}
  \label{fig:example_ackn}
\end{figure}

The choice of categories was inspired by \posscite{giles_who_2004} classification: funding agencies (FUND), corporations (COR), universities (UNI), and individuals (IND). For the present study, this classification was enhanced with the MISC (miscellaneous) and grant numbers (GRNB) categories. The GRNB category was adopted from WoS funding information indexing. The entities in the miscellaneous category could provide useful information but cannot be ascribed to other categories, e.g., names of projects and names of conferences. Figure~\ref{fig:example_ackn} demonstrates an example of acknowledged entities of different types. 

The Flair Named Entity Recognition (NER) model with Flair Embeddings\footnote{The model can be tested via an online demo \url{https://colab.research.google.com/drive/1Wz4ae5c65VDWanY3Vo-fj__bFjn-loL4?usp=sharing}} \citep{akbik_contextual_2018} was used to build a NER tagger \citep{smirnova_mayr_2022}. 
The Flair Embeddings model uses stacked embeddings, i.e., a combination of contextual string embeddings with a static embeddings model (in our case GloVe \citep{pennington_glove_2014}). Contextual string embeddings is a character based contextual string embeddings method proposed by \cite{akbik_contextual_2018}. This approach will generate different embeddings for the same word depending on its context. 
Our model was trained with a training corpus of 654 sentences extracted from the acknowledgement texts stored in WoS. Selection criteria for the training corpus were similar to the ones used for the acknowledgments corpus, i.e., records were restricted by year, scientific domain, article type, and language. The total accuracy\footnote{In our case accuracy computes as average of F1 scores of each entity type (\url{https://scikit-learn.org/stable/modules/generated/sklearn.metrics.f1_score.html}).} of the model is 0.77. The model is able to recognize six types of entities: FUND, GRNB, IND, COR, UNI and MISC. Some entity types (IND and GRNB) showed very good F1-Scores\footnote{Precision, Recall, and F1-Score are metrics to evaluate the model’s accuracy. Precision calculates the percentage of items that the model marked as positive that are actually positive, i.e., the percentage of labels assigned by the model that match the gold standard. The gold standard consists of a set of human annotations. Recall measures how many items that should have been marked as positive were actually marked as positive. The F1-Score is \textit{“a weighted harmonic mean of precision and recall”} (\url{https://web.stanford.edu/~jurafsky/slp3/4.pdf}).} over 0.9\footnote{F1-Score can take values from 0 to 1, where 1 indicates the perfect values of precision and recall.} \citep{smirnova_mayr_2022}. 

\paragraph{Data Disambiguation}\label{subsubsec:disamb}

After reviewing the first analysis of entities retrieved from the acknowledgements corpus, we realized that acknowledged entities need to be disambiguated for a more plausible analysis. Some entities have more than one writing variant as Example~\ref{ex:ex1} demonstrates. Ideally, these variants should be reduced to one representative. 

\begin{example}
\begin{itemize}
  \item National Science Foundation.
  \item NSF.
  \item National Science Foundation (NSF).
\end{itemize}
\label{ex:ex1}
\end{example}

The second problem that arose was that some different entities have identical abbreviations as Example~\ref{ex:ex2} demonstrates.

\begin{example}
\begin{itemize}
 \item Australia Awards Scholarship \hspace{1cm} AAS
 \item African Academy of Sciences \hspace{1cm} AAS
\end{itemize}
\label{ex:ex2}
\end{example}

Misspelling problems (Example~\ref{ex:ex3}) were existent for all entity types.

\begin{example}
\begin{itemize}
 \item National Nature Science Foundation of China
 \item Natural National Science Foundation of China
\end{itemize}
\label{ex:ex3}
\end{example}

To solve these problems, we created a disambiguation dataset containing frequent funding organisations and universities and their abbreviations. We analysed not-disambiguated data to find as much as possible variants of one entity, as Table~\ref{tab:disamb_pat} shows.

\begin{table}[!h]
\begin{center}
\caption{An example of the disambiguation dataset.}\label{tab:disamb_pat}%
\scalebox{0.8}{
\begin{tabular}{lll}
\toprule
                              text & abbreviation &                                     disambiguated form \\
\midrule
      Deutsches Klimarechenzentrum &         DKRZ &    German Climate Computer Center (DKRZ) \\
    German Climate Computer Center &         DKRZ &    German Climate Computer Center (DKRZ) \\
           German Computing Center &         DKRZ &    German Climate Computer Center (DKRZ) \\
                       UC Berkeley &          UCB & University of California, Berkeley (UCB) \\
University of California, Berkeley &          UCB & University of California, Berkeley (UCB) \\
\bottomrule
\end{tabular}
}
\end{center}
\end{table}

Furthermore, all entities in the model output dataset of FUND, UNI and MISC categories were compared to the entity names (column “text” in Tabel~\ref{tab:disamb_pat}), and abbreviations (column “abbreviation” in Table~\ref{tab:disamb_pat}) from the disambiguation datasets using the Levenshtein distance.  
The Levenshtein distance is a similarity measure that shows how similar two strings are\footnote{\url{https://en.wikipedia.org/wiki/Levenshtein_distance}}. We used the Python fuzz.ratio function\footnote{\url{https://github.com/seatgeek/thefuzz}} which calculates the Levenshtein distance similarity ratio between the two strings. 
The disambiguation corpus containing entities from the FUND category was used for the FUND and MISC categories. Meanwhile, the disambiguation corpus containing names of universities was used for the UNI category. For entity names, entries with the fuzz.ratio value more than 93 were replaced with the unified writing variant (the column “disambiguated form” in Tabel~\ref{tab:disamb_pat}). Thus, for entities in Example~\ref{ex:ex3}, it would be “National Science Foundation (NSF)”. For abbreviations, entries with fuzz.ratio value of more than 99 were considered as different variants of one abbreviation. As in this case, only entities with different upper- and lower-case writing should have been marked as one entity. Entities for which no matching patterns were found remained unchanged. 

The problem of various writing variants of the same entity also occurred in the COR category. In this case, a different approach was used. Tests showed that all variants of one entity could be found the most successfully using the fuzz.partial\_ratio function. The partial\_ratio picks the shortest string from the two compared strings and matches it with all substrings of the same lengths from the second string.  All the entities labeled COR were compared to each other using fuzz.partial\_ratio. Entries with a partial ratio value greater than 96 were identified as one entry. That fuzz.partial\_ratio value was determined by running tests on different writing variants of different COR entities. 

The problem with abbreviations that are similar for different entities was solved during the creation of the disambiguation dataset. Duplicated abbreviations were excluded from the disambiguation dataset. That way, if only the abbreviation (e.g., AAS) was in the output of the NER model (without its full name), then it would not be altered and would be placed in the original format into a disambiguated corpus.  

To solve the misspelling problem, all entities were compared to each other within their entity types (e.g., FUND with FUND, MISC with MISC, etc.) using Levenshtein distance. Entities with the Levenshtein distance of more than 90 were identified as the same entity. For the IND category, entities with the Levenshtein distance equal to 100 were identified as one category; as in this case, only entities which differ only in upper- and lower-case writing variants (e.g.,\textit{ John Doe} vs.\textit{ john doe}) were considered as different writing variants of the same entity.

Additionally, the top 30 most frequent entities were manually reviewed. Entities mistakenly classified to the wrong categories were placed to the appropriate category in the whole dataset \footnote{For example entity World Bank in economics was distributed between COR and FUND categories. We moved all mentions of that category from the FUND category into the COR category.}. Furthermore, the IND category had some entities like \textit{Drs.} or \textit{J.}, which do not carry meaningful information, as Table~\ref{tab:ind_notdis_example} demonstrates. For that reason, individuals containing less than 4 characters (with spaces) were removed from the dataset. The same procedure was conducted with the grant numbers.   


\begin{table}[!h]
\begin{center}
\caption{An example of disambiguated but not manually reviewed entities from the IND category.}\label{tab:ind_notdis_example}%
\scalebox{0.6}{
\begin{tabular}{llllllll}
\toprule
   oceanography &  &        social sciences & &        economics &  & computer science &  \\
\midrule
            IND &       frequency &                    IND &       frequency &              IND &       frequency &              IND &       frequency \\
\midrule
           Drs. &         89 & Kathleen Mullan Harris &        289 &    Jan Wallander &        163 &         H. Zhang &         27 \\
             J. &         72 &           Richard Udry &        158 &  Andrei Shleifer &        115 &            X. Li &         26 \\
   Kelly Cooper &         68 &       Peter S. Bearman &        149 &               Li &         89 &               J. &         25 \\
     Ki-Han Kim &         45 &       Barbara Entwisle &        120 &       Raj Chetty &         85 &           P. Shi &         23 \\
Maria S. Merian &         44 &     Ronald R. Rindfuss &        117 & Sushanta Mallick &         83 &           J. Liu &         23 \\
             M. &         44 &                John D. &         72 &   Andrew Karolyi &         83 &          H. Wang &         23 \\
             R. &         35 &          Oliver Linton &         60 &   Iftekhar Hasan &         80 &          L. Wang &         22 \\
\bottomrule
\end{tabular}
}
\end{center}
\end{table}

\section{Results}\label{sec:res}

\begin{figure}[h!]
\centering
  \includegraphics[width=1\textwidth]{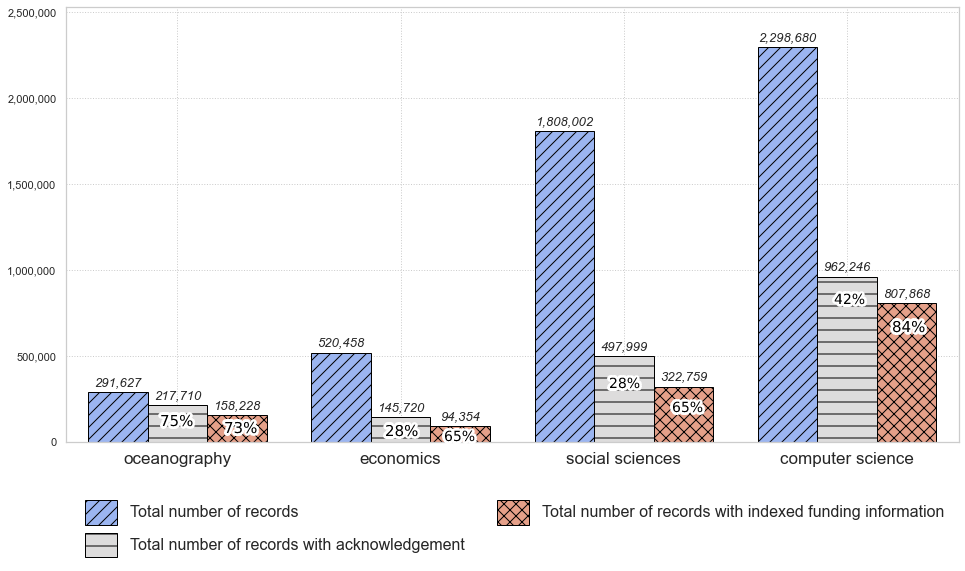}
  \caption{Analysis of WoS entries from four scientific domains published between 2014 and 2019 (publications in English).}
  \label{fig:analysis_tot_ackn}
\end{figure}

Prior to extracting data for the acknowledgement corpus, we performed a preliminary analysis of the entries stored in WoS \footnote{The analysis was performed on the disambiguated (cleaned) data.}. Only articles from four scientific domains published between 2014 and 2019 in English were considered. As Figure~\ref{fig:analysis_tot_ackn} demonstrates, not every WoS entry has an acknowledgement text. This corresponds to the findings of \cite{paul_hus_all_2016}. Thus, oceanography has the smallest amount of articles, but at the same time is the best covered area in terms of availability of acknowledgement texts: 75~\% of all articles posses an acknowledgement text. Moreover 73~\% of these articles have indexed funding information. The computer science domain includes the largest amount of entries, 42~\% of which are in possession of an acknowledgment text. At the same time 84~\% of these articles have an indexed funding information, which makes the computer science domain to the best covered domain in terms of funding information indexing. In both economics and social science only 28~\% of all records include acknowledgement text and only 65~\% of these records have indexed funding information. 
 
\subsection{Acknowledgment Trends }\label{subsec:ackn_trends}

\begin{figure}[h!]
\centering
  \includegraphics[width=1\textwidth]{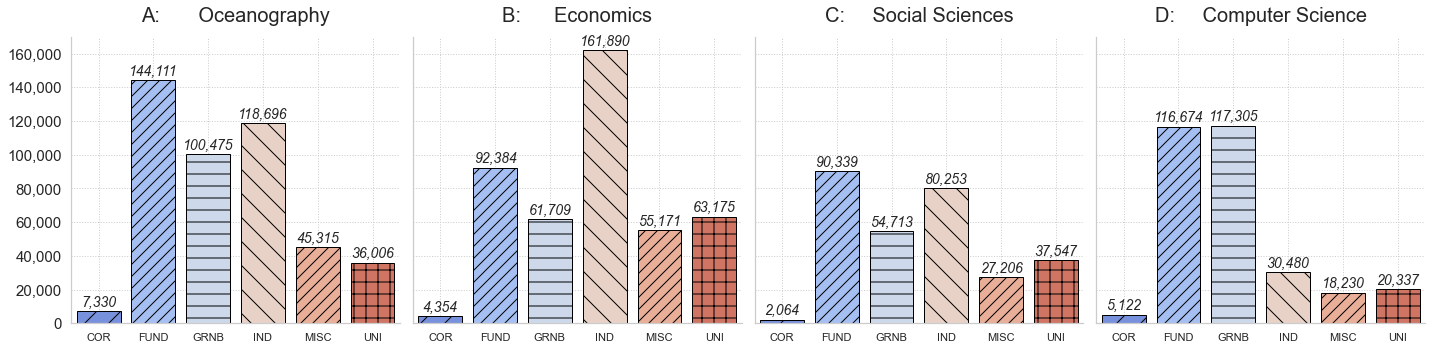}
  \caption{The distribution of acknowledged entities between disciplines.}
  \label{fig:labels_freq}
\end{figure}

\begin{figure}[h!]
\centering
  \includegraphics[width=1\textwidth]{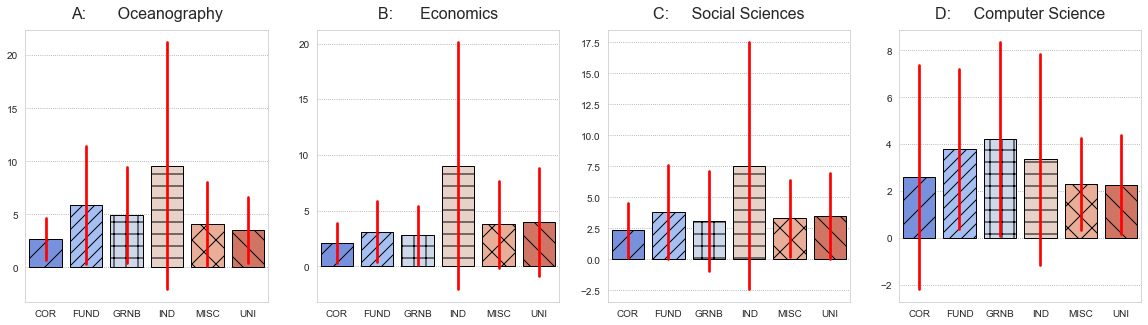}
  \caption{Mean number of acknowledged entities (with the standard deviation) per paper in different disciplines. Note: the scale of the y-axis is non-uniform due to clearness.}
  \label{fig:labels_mean_std}
\end{figure}

Figure~\ref{fig:labels_freq} shows the distribution of entities of different types between scientific domains. The distribution of entities demonstrates clear differences among scientific domains. Therefore, IND is the most frequent entity type in economics, while oceanography demonstrates a highest mean number of acknowledged individuals per acknowledgement text among the highest standard deviation (as Figure~\ref{fig:labels_mean_std}-A and Table~\ref{app:mean_std} demonstrate), which means that the numbers of acknowledged individuals in some acknowledgement texts are significantly more or less as mean values. In general, all scientific domains (Figure~\ref{fig:labels_mean_std}) show the highest standard deviation for the numbers of acknowledged individuals. In computer science as opposed to other disciplines, the highest mean number of acknowledged entities belongs to grant numbers. Overall, all disciplines showed high dispersion of number of all acknowledged entities relating to the mean values.  
FUND is the most frequent in social science and oceanography; at the same time, GRNB is the most frequent entity in computer science. However, social science and oceanography domains show similar frequency patterns of FUND, IND and GRNB categories: FUND is the most frequent category followed by IND and GRNB. COR is the most under-represented category in all scientific domains. UNI and MISC have similar distributions in economics, social sciences and computer science: more entities fall into the UNI category than into the MISC category. On the other hand, oceanography shows the opposite pattern. From all scientific domains, computer science demonstrates the lowest number of acknowledged individuals and the highest number of grant numbers.

\begin{figure}[h!]
\centering
  \includegraphics[width=0.8\textwidth]{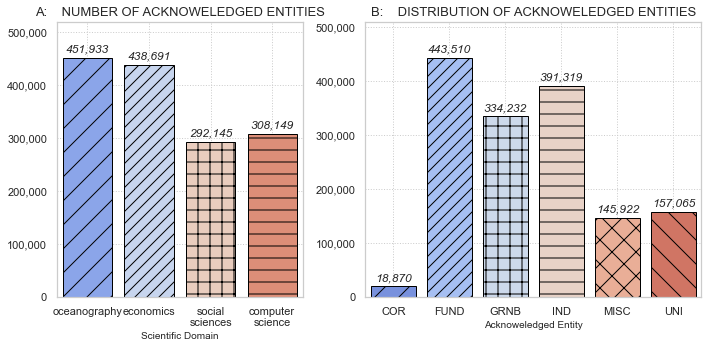}
  \caption{The general distribution of acknowledged entities categories. Figure A represents the total amount of acknowledged entities among disciplines and Figure B represents the general distribution of acknowledged entities.}
  \label{fig:ackn_ent_counts}
\end{figure}

Overall, oceanography showed the highest number of acknowledged entities, followed by economics, computer science and social sciences, as Figure~\ref{fig:ackn_ent_counts}-A demonstrates. Oceanography and economics, and social sciences and computer science have minor differences between the numbers of acknowledged entities, while there is a bigger gap between the two pairs. As Figure~\ref{fig:ackn_ent_counts}-B shows, FUND is the most frequent entity followed by IND, GRNB, UNI, MISC and COR. IND and GRNB are the second most frequent categories. The COR category has the lowest number of occurrences.

\begin{figure}[h!]
\centering
  \includegraphics[width=0.8\textwidth]{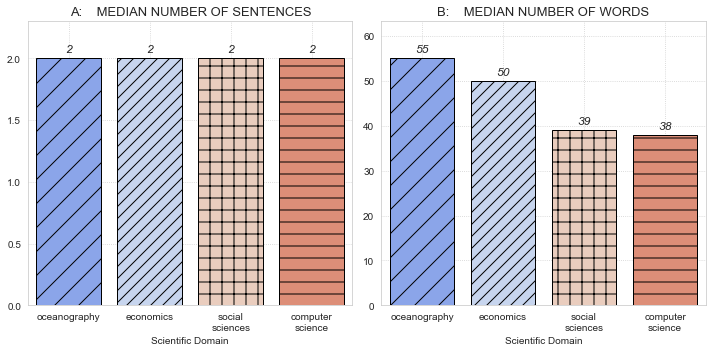}
  \caption{Median number of sentences (Figure A) and median number of words (Figure B) in acknowledgment texts in different scientific domains.}
  \label{fig:sent_word_counts_bar}
\end{figure}

Analysis of the lengths of acknowledgement texts showed that in general in all four scientific domains an acknowledgement consists of two sentences, which comprise 38 (computer science), 39 (social sciences), 50 (economics), and 55 (oceanography) words\footnote{Punctuation marks were excluded from the count.}. Thus, as Figure~\ref{fig:sent_word_counts_bar} demonstrates, the oceanography domain has in general longer acknowledgement texts with longer sentences followed by economics, social sciences and computer science. 

\begin{figure}[h!]
\centering
  \includegraphics[width=1\textwidth]{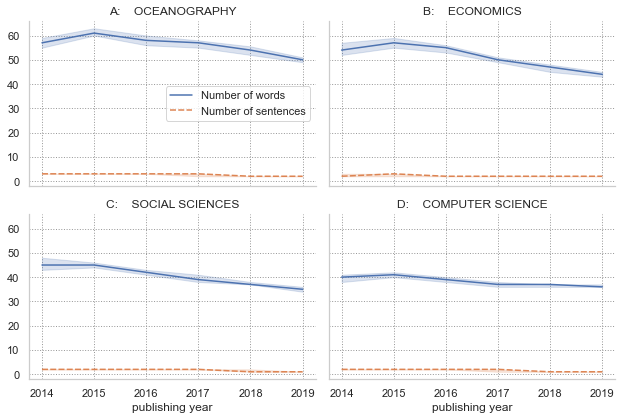}
  \caption{Development of the lengths of the acknowledgement texts from four scientific domains in 5 years (2014 - 2019). The figure shows median and 95\% confidence interval for each unit. }
  \label{fig:sent_word_counts}
\end{figure}

Figure~\ref{fig:sent_word_counts} depicts the development of the lengths of acknowledgement texts in four scientific domains from 2014 till 2019. Changes in the number of sentences were observed for all scientific domains. Thus, in economics(Figure~\ref{fig:sent_word_counts}-B) the number of sentences increased from 2 in 2014 to 3 in 2015 and then decreased to 2 again in the following years. In social sciences (Figure~\ref{fig:sent_word_counts}-C), oceanography (Figure~\ref{fig:sent_word_counts}-A) and computer science (Figure~\ref{fig:sent_word_counts}-D) the similar pattern can be observed: the number of sentences decreased by 50\% in computer science and social sciences and by 33\% in oceanography starting from the year 2018. 
Number of words also decreased in all scientific domains. Thus, the highest percentage decrease of 22\% showed social sciences, followed by economics (19\%) and oceanography (12\%). The lowest value of decrease of 10\% is in the computer science domain.
The highest range in the median number of words was observed in economics and social sciences in 2014. The smallest amount of variations showed computer science. In general, from 2014 to 2019 the value range decreased in all disciplines.

\begin{figure}[h!]
\centering
  \includegraphics[width=1\textwidth]{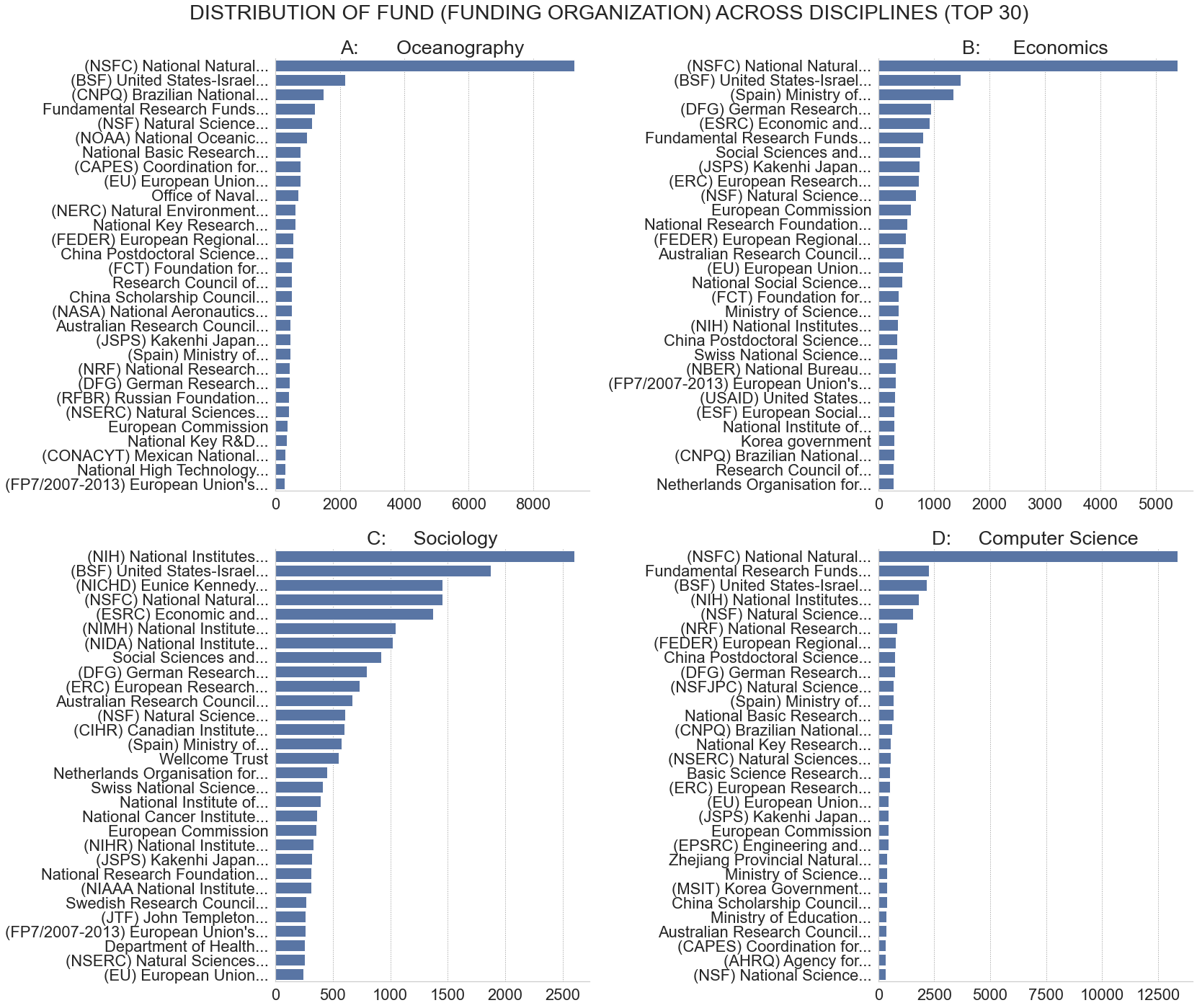}
  \caption{Top 30 acknowledged entities which fall into the FUND (funding organization) category. Figure A represents entities from oceanography, Figure B from economics, Figure C from social science, and Figure D from computer science.  }
  \label{fig:fund}
\end{figure}

\begin{figure}[h!]
\centering
  \includegraphics[width=1\textwidth]{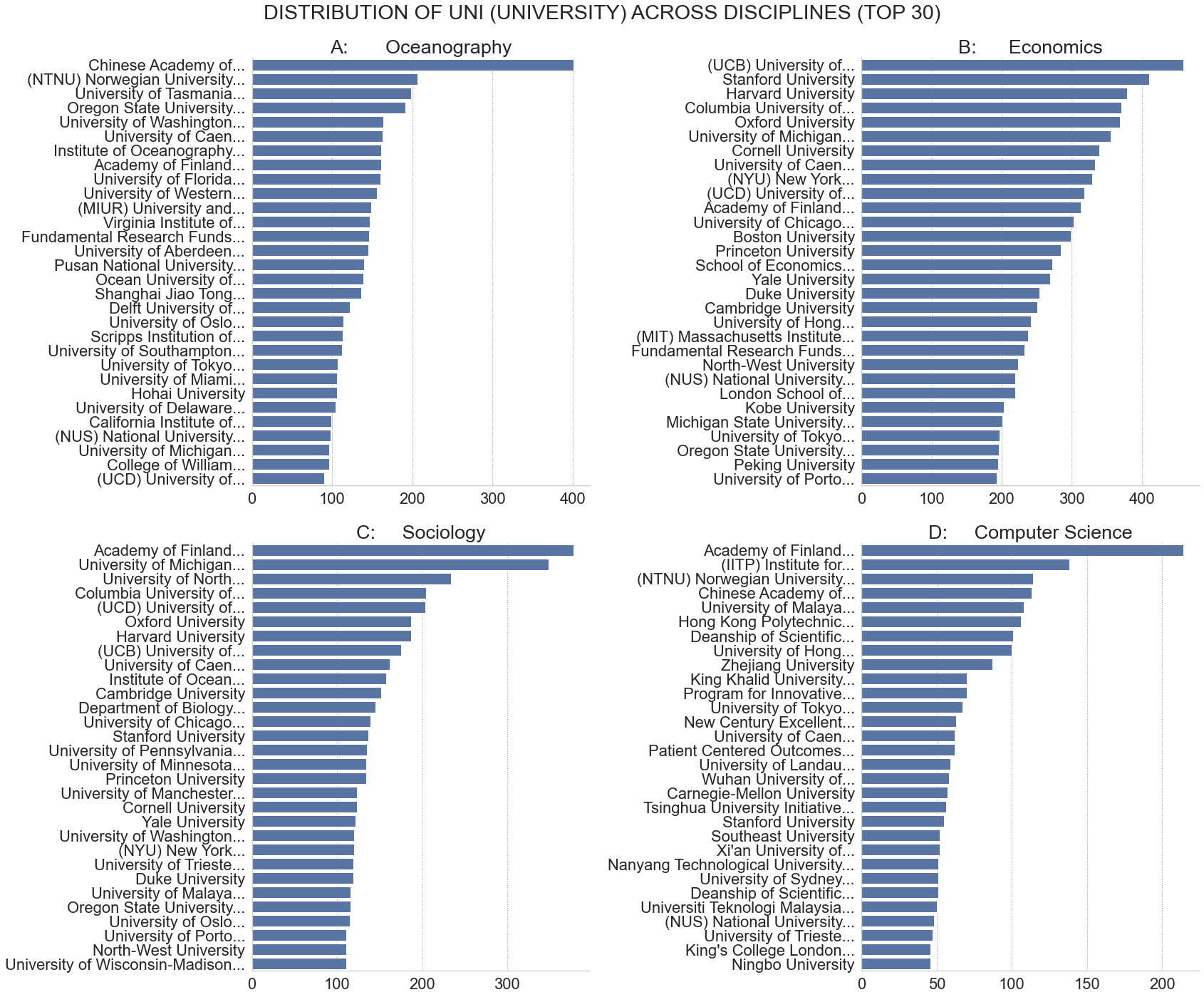}
  \caption{Top 30 acknowledged entities which fall into the UNI (universities) category. Figure A represents entities from oceanography, Figure B from economics, Figure C from social science, and Figure D from computer science.  }
  \label{fig:uni}
\end{figure}

\begin{figure}[h!]
\centering
  \includegraphics[width=1\textwidth]{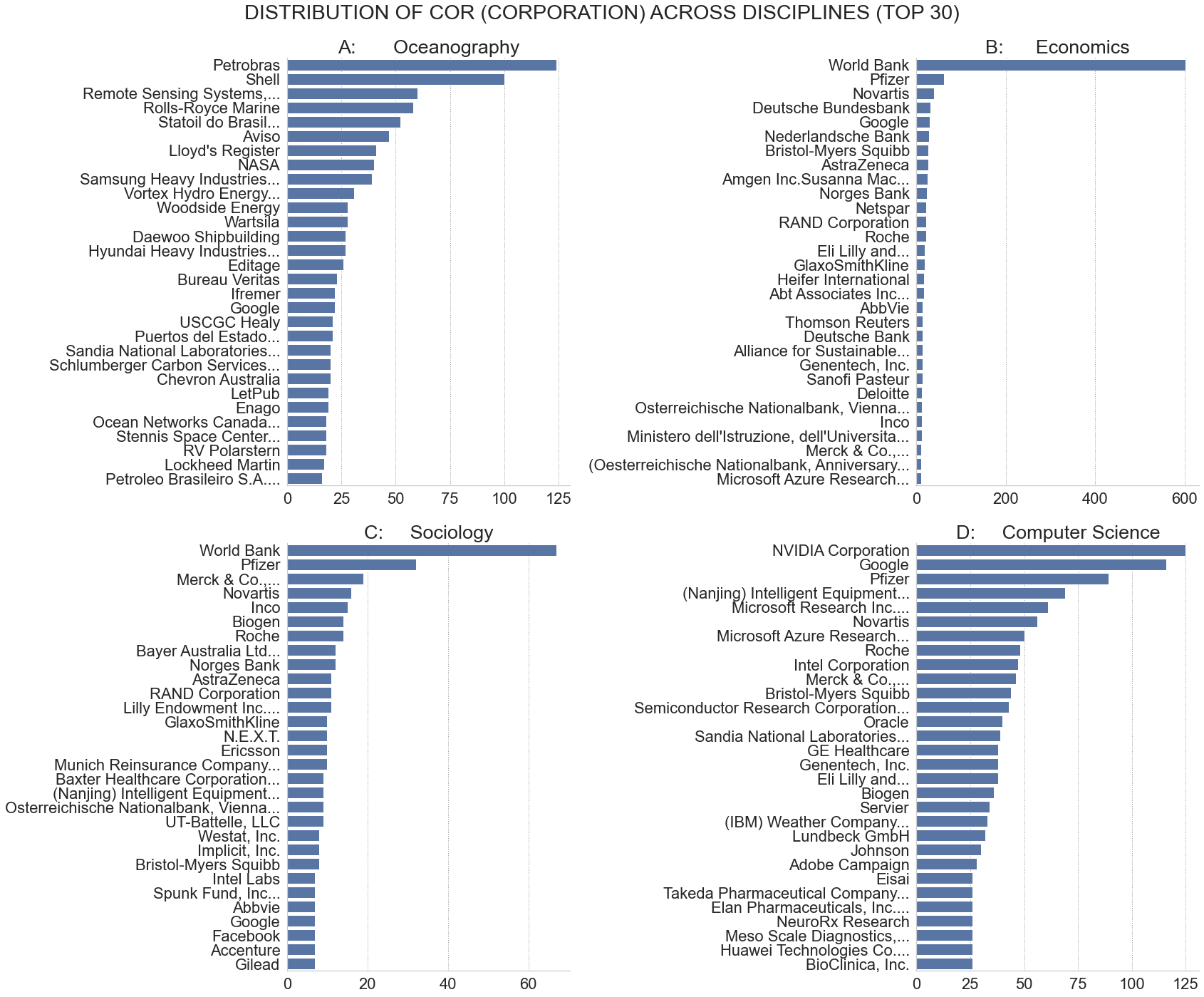}
  \caption{Top 30 acknowledged entities which fall into the COR (corporation) category. Figure A represents entities from oceanography, Figure B from economics, Figure C from social science, and Figure D from computer science.  }
  \label{fig:cor}
\end{figure}

Figures~\ref{fig:fund},\ref{fig:uni},\ref{fig:cor},\ref{fig:ind}, and \ref{fig:misc} demonstrate the top 30 acknowledged entities of different types except GRNB. Some interesting trends can be observed as described further below. 

Figure~\ref{fig:fund} shows the top 30 most frequent entities that fall under the FUND category.  All scientific domains, except social sciences, have similar top-first funding organizations: the National Natural Science Foundation of China (NSFC), whereas the top-first funding organization for social sciences is the National Institutes of Health (NIH).  The United States-Israel Binational Science Foundation (BSF) is the second most frequent funding organization for all disciplines, except computer science, which second most frequent funding organization is Fundamental Research Funds for the Central Universities and the United States-Israel Binational Science Foundation (BSF) took the third place. Mainly, the major funders such as the National Natural Science Foundation of China, National Institutes of Health, Deutsche Forschungsgemeinschaft (DFG), etc. can be found in the top 30 funding organizations list. 

Figure~\ref{fig:uni} shows the top 30 most frequent entities which fall into the UNI category. The Academy of Finland is the first top university for social sciences and computer science. The Norwegian University of Science and Technology (NTNU) is in the second position for oceanography and in the third position for computer science. The University of California, Berkeley (UCB) is in the top position for economics and the Chinese Academy of Sciences for oceanography. 

Figure~\ref{fig:cor} depicts the top 30 most frequent entities which fall into the COR category and entities from this category tend to be scientific domain specific. Thus, the top two most frequent corporations in computer science (NVIDIA Corporation and Google) are companies from the software and computing field. The top first and third entities from oceanography (Petrobras and Shell) are companies from the petroleum industries. Pfizer (a pharmaceutical and biotechnology corporation) made the top third in economics, computer science and social sciences. The top second for social sciences and the top third for economics is Novartis, a pharmaceutical company. The World Bank is the most frequent entity for economics and social sciences. The top third for social sciences is Merck \& Co., Inc, which is also a pharmaceutical corporation. 

Overall, the distribution of acknowledged entities in oceanography, economics and computer science has the following pattern: the top first most frequent entity has the greatest number of occurrences, other entities starting from the top second occur massively less frequently. On the other hand, the distribution of entities in social sciences is more uniform. That distribution pattern can be observed (more or less defined) almost for all entity types in all scientific domains and correspond with the \posscite{giles_who_2004} findings and follows a power law, which postulates that\textit{ "few entities are named very frequently and a great many entities are named only rarely"} \citep[~p.17601]{giles_who_2004}. Nevertheless, it was observed that some entities do not follow the power law mentioned above. This applies to the entities of UNI category from economics domain (Figure~\ref{fig:uni}-B) and IND category from computer science domain (Figure~\ref{fig:ind}-D).

\subsection{Relationships between analysed variables.}\label{subsec:correllation}

A Chi-square test of independence was conducted to examine the relationship between an entity, entity type and scientific domain. 
As Table~\ref{tab:chi2} demonstrates, the relation between these variables was significant. The P-value of all pairs of examined variables tends toward zero. 

\begin{table}[h!]
\begin{center}
\caption{The results of the Chi-square test for multiple variables. The column “Variables” represents variable pairs that were examined.}\label{tab:chi2}%
\begin{tabular}{lrrr}
\toprule
       variables &  Chi-Square &  P-value &  degrees of freedom \\
\midrule
 entity type - scientific field &  143606.086 &      0.0 &                  15 \\
entity - scientific field & 3264463.167 &      0.0 &             1343037 \\
    entity - entity type & 7107380.717 &      0.0 &             2238395 \\
\bottomrule
\end{tabular}
\end{center}
\end{table}

However, a weak association in a large sample size may also result in a very low P-value\footnote{\url{https://www.spss-tutorials.com/cramers-v-what-and-why/}}. To assess the strengths of association between an acknowledged entity, entity type and a scientific domain, the Cramer V test was conducted. 
The results of the analysis are represented in Figure~\ref{fig:cramer}. Analysis was performed on the disambiguated data (see Section~\ref{subsubsec:disamb}). Table~\ref{tab:dof} demonstrates the degrees of freedom for the analyzed variable pairs. 

A large association indicates that by knowing the value of one category, the value of a second related category can be predicted. No association, on the other hand, indicates that the categories are not related. As expected, there is a large association between the acknowledged entity and the type of entity (Cramer's V is equal to 0.98). The variable pair entity type and scientific domain showed a small association of 0.18. Furthermore, a large association between the entity and scientific domain was discovered (0.85). Thus, a precise acknowledged entity would, with high probability, have a precise label and would belong to the precise scientific domain.

\begin{table}[h!]
	\begin{minipage}{0.55\linewidth}
		\centering
		\includegraphics[width=0.9\textwidth]{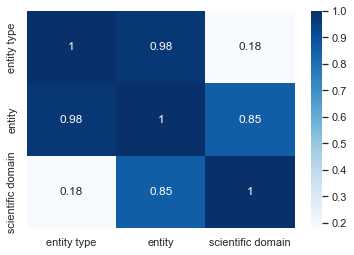}
		\captionof{figure}{Distribution of the Cramer’s V values between scientific domains, entity types and acknowledged entities.}\label{fig:cramer}
	\end{minipage}\hfill
	\begin{minipage}{0.4\linewidth}
	\caption{Degrees of freedom for the variable pairs.}
		\label{tab:dof}
		\centering
			\resizebox{\textwidth}{!}{%
		\begin{tabular}[width=\linewidth]{@{}lll@{}}
    \toprule
    \ Variable & Entity type & Scientific domain \\  
    \midrule
			Entity type &  & 3 \\
            Entity & 5 & 3 \\
    \bottomrule
    \end{tabular}}
    \end{minipage}
\end{table}

\begin{figure}[h!]
\centering
  \includegraphics[width=0.8\textwidth]{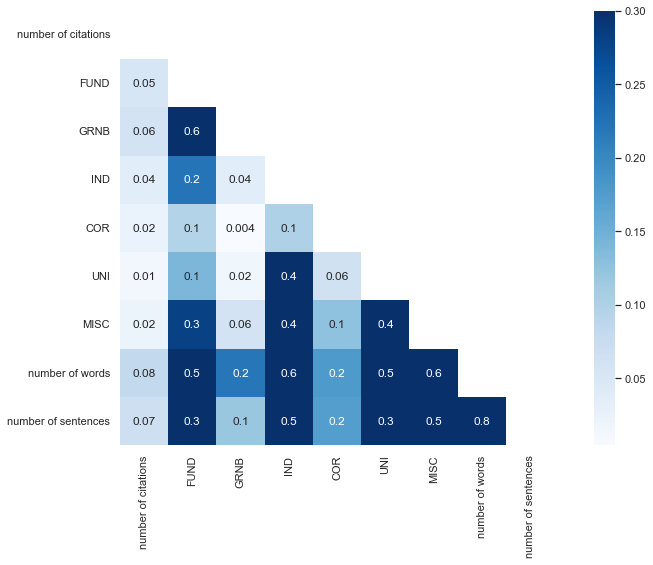}
  \caption{Correlation between number of citations of the article, number of acknowledged entities of different types, number of sentences and number of words in the acknowledgement text.}
  \label{fig:correl_all}
\end{figure}

Relationships between different numerical variables (number of citations, number of acknowledged entities, number of words, and number of sentences in the acknowledgement texts) were examined with the use of the Pearson correlation coefficient (Pearson's R) \footnote{The analysis was performed on the normalized data.}.
As Figure~\ref{fig:correl_all} demonstrates, no correlation between number of citations and number of acknowledged entities was found: values of the Pearson's R for all variable pairs is under 0.1. Furthermore, no correlation was found between the number of citations and number of words and sentences in the acknowledgement text.

High correlation was observed between number of words and number of acknowledged funding organisations (0.5), individuals (0.6), universities (0.5) and miscellaneous entities (0.6). The number of sentences showed a high correlation with the number of acknowledged individuals (0.5) and miscellaneous entities (0.5) and a moderate correlation with the number of acknowledged funding organisations (0.3) and universities (0.3). Clearly, the number of sentences in the acknowledgement text is highly correlated with the number of words (0.8). 

Additionally, a high correlation (0.6) was discovered between the number of grant numbers and number of acknowledges funding organisations. A moderate correlation of 0.4 was observed between the number of acknowledged universities and the number of individuals and miscellaneous entities along with between the number of acknowledged individuals and miscellaneous entities. 

The ANOVA test was performed on the analysed variables. The P-value is very low (0.000), which means that the results are statistically significant.

\section{Discussion}\label{sec:disc}
As we expected, the relations between entity types and scientific domains were found. A strong association was found between acknowledged entity and scientific domain, and acknowledged entity and entity type, as we anticipated. Nevertheless, an association between scientific domain and entity type was minor. Furthermore, social sciences was expected to have the highest number of acknowledged individuals (as in \cite{paul_hus_all_2017}) and economics (as in \cite{diaz-faes_making_2017}), the results partly coincided with our expectation: our analysis showed that individuals were most acknowledged in economics. Social sciences and computer science in general showed a smaller number of acknowledged entities than economics and oceanography. This can possibly be explained by the fact that the acknowledgement corpus generally contains fewer acknowledgements from these two scientific domains than from economics and oceanography (Table~\ref{tab:ackn_corpus_count}). 

The prevalence of the FUND category is not surprising, as WoS stores only acknowledgements containing funding information. Unexpectedly, the IND category was the second most frequent category: GRNB was expected to be one of the most frequent categories.
Interesting trends were observed in the COR category. The most research in social sciences and economic domains is funded by pharmaceutical corporations and banks, while the top corporations acknowledged in the oceanography domain come from the petroleum industry. The top corporations in computer science are in the software and computing field.  


Generally, the number of words in the acknowledgement texts positively correlates with the number of acknowledged funding organizations, universities, individuals and miscellaneous entities. At the same time, acknowledgement texts with the larger number of sentences have more acknowledged individuals and miscellaneous categories. Correlations between frequencies of entities of different types were found. Thus, acknowledgement texts with more acknowledged funding organizations would posses more grant numbers. Acknowledgements with more acknowledged individuals will have more acknowledged universities and miscellaneous entities and vice versa. 

Our analysis revealed that oceanography in general have the longest acknowledgement texts, while computer sciences domain the shortest ones. Generally number of words in an acknowledgement text decreased from 2014 to 2015 in all scientific domains.

\posscite{giles_who_2004} analysed the correlations between the number of acknowledgements received by an individual and that individual’s h-index. H-index is an index used to assess an individual’s scientific research output \citep{hirsch_2005}.  \cite{giles_who_2004} used computer science research papers from the CiteSeer digital library as a data source. We tried to replicate the \posscite{giles_who_2004} study, but faced difficulties in referencing an acknowledged individual name and a real person\footnote{For example, it is yet impossible to detect individuals, which were acknowledged with abbreviated names. Some identical names can belong to different people.}. To solve this problem, additional disambiguation techniques should be developed. Many acknowledgement texts contain name of a person together with the person's institutional affiliation (usually names of universities or corporations), therefore it is possible to identify the acknowledged person through the name of person's affiliation. The Flair framework additionally provides the position of the acknowledged entity in the text (i.e. index range of the characters). Thus, an acknowledged university or corporation, which is in the close index range with the acknowledged individual would highly likely be the the person's institutional affiliation.

Additionally, we would propose to include an ORCID of the acknowledged individual while writing an acknowledgement text, to make the identification procedure easier.

\section{Conclusion}\label{sec:conc}

The analysis of the automatically extracted entities revealed differences and distinct patterns in the distribution of acknowledged entities of different types between different scientific domains. The extracted funding information poses various possibilities and challenges for further research. Thus, grant numbers, funding organizations and corporations might give an insight on the impact of funding sponsorship on scientific research. Names of individuals might assess one's scholarly performance, and the miscellaneous category might reveal other non-sponsorship variables that might influence academic research. 

One of the future research aims might be the analysis of correlations between the acknowledged entities and such characteristics as journal impact, other scientific domains, researcher’s productivity and impact. Furthermore, some funding organizations might find the extracted acknowledgements information useful in order to track if funding recipients acknowledge funding in their publications. The disambiguation of recognised entities poses further interesting challenges, since in the present paper only a superficial disambiguation of the data was conducted which was necessary for the analysis and visualization of the results. 

The main limitation of our study was the peculiarities of the WoS funding information indexing. The WoS includes only acknowledgements that contain funding information; therefore, not every database entry has an acknowledgement. This can explain the prevalence of funding agencies over other types of entity.  

\section{Acknowledgement}

This work was funded by German Centre for Higher Education Research and Science Studies (DZHW) via the project "Mining Acknowledgement Texts in Web of Science (MinAck)"\footnote{\url{https://kalawinka.github.io/minack/}}. 
Nina Smirnova acknowledges support by Deutsche Forschungsgemeinschaft (DFG) under grant  number MA 3964/7-2, the Fachinformationsdienst Politikwissenschaft -- Pollux.
Access to the WoS data was granted via the Competence Centre for Bibliometrics\footnote{\url{https://www.bibliometrie.info/en/index.php?id=home }}. Data access was funded by BMBF (Federal Ministry of Education and Research, Germany) under grant number 01PQ17001.

\clearpage
\begin{appendices}
\section{List of WoS disciplines used for the present paper}\label{app:wos_disc}
\begin{tabular}{ll}
\toprule
                                  CLASSIFICATION &        SCIENTIFIC FIELD \\
\midrule
                              Engineering, Ocean &     oceanography \\
                     Marine \& Freshwater Biology &     oceanography \\
                                    Oceanography &     oceanography \\
                             Engineering, Marine &     oceanography \\
                                       Economics &        economics \\
                 Agricultural Economics \& Policy &        economics \\
                            Business \& Economics &        economics \\
                      Biomedical Social Sciences &  social sciences \\
                                     Social Work &  social sciences \\
                     Social Sciences, Biomedical &  social sciences \\
              Social Sciences, Interdisciplinary &  social sciences \\
                  Social Sciences - Other Topics &  social sciences \\
                                   Social Issues &  social sciences \\
                              Psychology, Social &  social sciences \\
           Social Sciences, Mathematical Methods &  social sciences \\
                                       Sociology &  social sciences \\
                      History Of Social Sciences &  social sciences \\
                                 Social Sciences &  social sciences \\
         Mathematical Methods In Social Sciences &  social sciences \\
          Computer Science, Software Engineering & computer science \\
       Computer Science, Hardware \& Architecture & computer science \\
       Computer Science, Artificial Intelligence & computer science \\
                                Computer Science & computer science \\
                             Medical Informatics & computer science \\
           Computer Science, Information Systems & computer science \\
Computer Science, Interdisciplinary Applications & computer science \\
                   Computer Science, Cybernetics & computer science \\
              Computer Science, Theory \& Methods & computer science \\
\bottomrule
\end{tabular}

\clearpage
\section{Top 30 acknowledged individuals and miscellaneous entities}\label{app:ackn_ind}

\begin{figure}[h!]
\centering
  \includegraphics[width=1\textwidth]{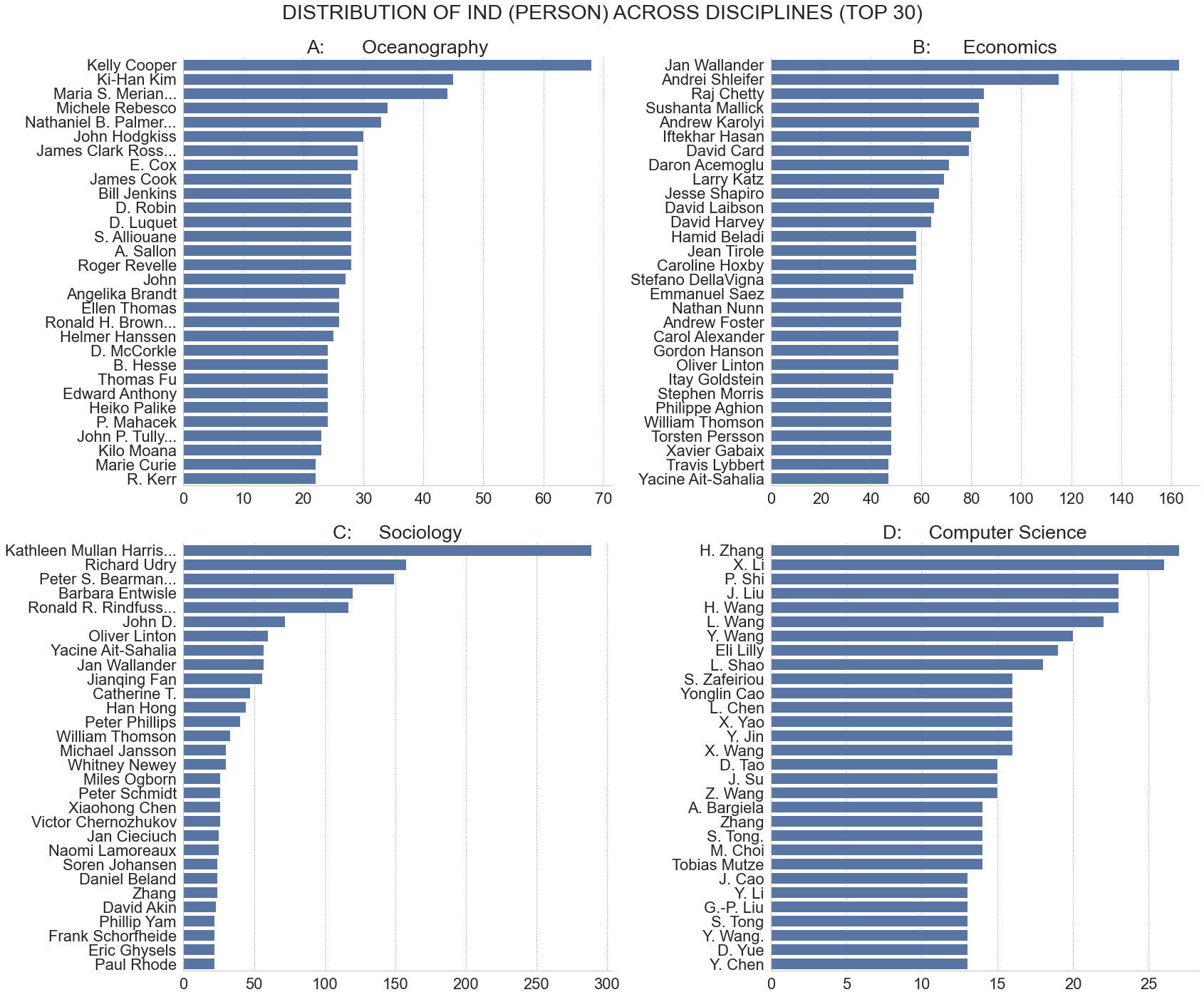}
  \caption{Top 30 acknowledged entities which fall into the IND (person) category. Figure A represents entities from oceanography, Figure B from economics, Figure C from social science, and Figure D from computer science. }
  \label{fig:ind}
\end{figure}

\begin{figure}[h!]
\centering
  \includegraphics[width=1\textwidth]{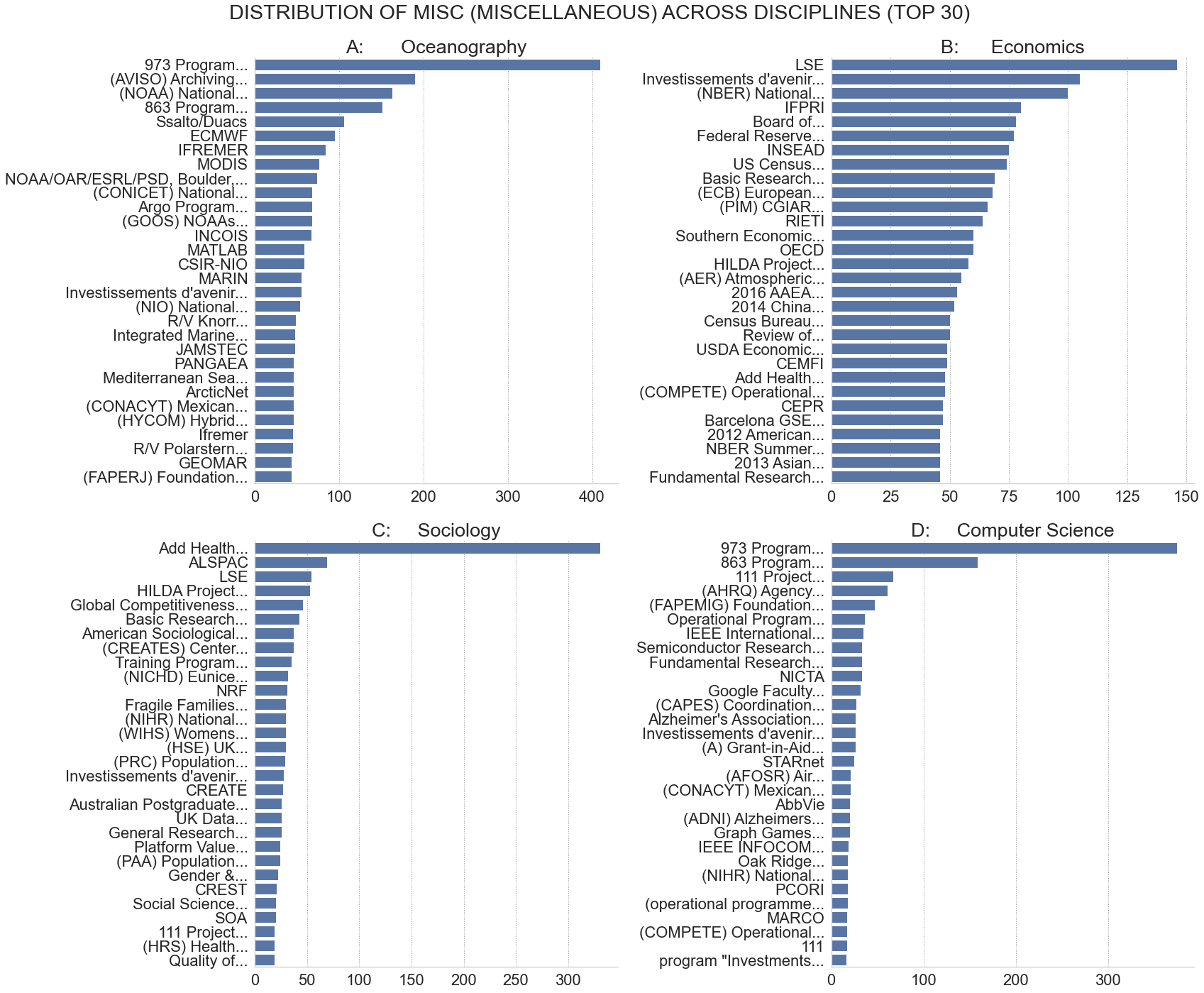}
  \caption{Top 30 acknowledged entities which fall into the MISC (miscellaneous) category. Figure A represents entities from oceanography, Figure B from economics, Figure C from social science, and Figure D from computer science.}
  \label{fig:misc}
\end{figure}
\clearpage

\section{Mean number of acknowledged entities per paper in different disciplines including standard deviation.}\label{app:mean_std}
\begin{tabular}{llrr}
\toprule
      Scientific field & Label &     Mean &       Standard deviation \\
\midrule
computer science &   COR & 2.601320 &  4.790851 \\
computer science &  FUND & 3.810634 &  3.411334 \\
computer science &  GRNB & 4.213542 &  4.129411 \\
computer science &   IND & 3.360900 &  4.488084 \\
computer science &  MISC & 2.302349 &  1.951288 \\
computer science &   UNI & 2.275596 &  2.118370 \\
       economics &   COR & 2.103382 &  1.808882 \\
       economics &  FUND & 3.125622 &  2.751300 \\
       economics &  GRNB & 2.778183 &  2.649553 \\
       economics &   IND & 9.017434 & 11.094021 \\
       economics &  MISC & 3.773150 &  3.924555 \\
       economics &   UNI & 4.002978 &  4.856648 \\
    oceanography &   COR & 2.633848 &  1.977578 \\
    oceanography &  FUND & 5.871298 &  5.543036 \\
    oceanography &  GRNB & 4.906725 &  4.525458 \\
    oceanography &   IND & 9.569943 & 11.648552 \\
    oceanography &  MISC & 4.067774 &  3.938683 \\
    oceanography &   UNI & 3.547739 &  3.105968 \\
 social sciences &   COR & 2.332203 &  2.235140 \\
 social sciences &  FUND & 3.807435 &  3.771419 \\
 social sciences &  GRNB & 3.118261 &  4.026767 \\
 social sciences &   IND & 7.541158 &  9.967386 \\
 social sciences &  MISC & 3.324291 &  3.095832 \\
 social sciences &   UNI & 3.506444 &  3.447353 \\
\bottomrule
\end{tabular}

\end{appendices}

\clearpage
\typeout{} 
\bibliography{sn-bibliography}



\end{document}